\documentclass[twocolumn]{jpsj3}
\usepackage{txfonts}
\graphicspath{{fig/}}

\title{
Vortex State and 
Field-Angle Resolved Specific Heat Oscillation for $H \parallel ab$ 
in d-Wave Superconductors
}

\author{
Masayuki Hiragi,
Kenta M. Suzuki~\thanks{E-mail address: kenta@mp.okayama-u.ac.jp},  
Masanori Ichioka~\thanks{E-mail address: oka@mp.okayama-u.ac.jp}, 
Kazushige Machida~\thanks{E-mail address: machida@mp.okayama-u.ac.jp}
}
\inst{
Department of Physics, Okayama University,
Okayama 700-8530, Japan
}
\abst{
When magnetic field is applied parallel to the $ab$ plane 
in $d_{x^2-y^2}$-wave superconductors, 
the transition of stable vortex lattice structure, 
spatial structure of local density of states, 
and specific heat oscillation by rotation of magnetic field orientation 
are investigated by quantitative calculations based on the  selfconsistent Eilenberger theory. 
We estimate how the vortex state changes depending on the relative angle between 
the node-direction of the superconducting gap and magnetic field orientation. 
To reproduce the sign-change of specific heat oscillation observed in 
${\rm CeCoIn_5}$, our study is done by including strong paramagnetic effect. 
The quantitative theoretical calculations give decisive information 
to analyze the experimental data on the field-angle dependence,
and establish the angle-resolved specific heat experiment as
a spectroscopic means to identify the node-position of the superconducting gap. 
}

\kword{
vortex state under parallel field, 
paramagnetic effect,  
specific heat oscillation, 
Eilenberger theory
}

\begin{document}
\maketitle

\def\runtitle{Field-angle resolved heat capacity oscillation and gap symmetry in d-wave superconductors}
\def\runauthor{M.Hiragi, {\it et al.} }
\section{Introduction}

In unconventional superconductors such as heavy fermion superconductors, 
high-$T_{\rm c}$ cuprate superconductors, organic superconductors, 
the pairing mechanism may not be conventional electron-phonon interaction.  
The Cooper pairs are formed by exotic pairing mechanism 
coming from the strong electron-electron interaction,  
and possibly the pairing symmetry is $p$-, $d$-, or $f$-wave, 
other than full gap $s$-wave. 
Thus, the pairing function has sign change on the Fermi surface, 
resulting in nodes of the superconducting gap.   
To clarify the mechanism of unconventional superconductors, 
it is important to identify the pairing symmetry of the Cooper pairs. 

Experimentally, 
existence of the node can be detected by the power-law dependence 
of physical quantities as a function of temperature $T$.  
For example, the specific heat $C \propto T^2$ ($\propto T^3$) 
and nuclear spin relaxation rate $T_1^{-1} \propto T^3$ ($\propto T^5$)
for line (point) node of superconducting gap, instead of exponential 
$T$-dependence for full gap $s$-wave pairing~\cite{sigrist}. 
This comes from the power-law of the density of states (DOS) 
$N(E) \propto E$  ($\propto E^2$) for the line (point) nodes of 
the gap function. 

The position of the node on the Fermi surface 
can be identified by the phase-sensitive experimental methods such 
as corner junction~\cite{wollman,tsuei}   
and Andreev surface bound states~\cite{kashiwaya}. 
For these experiments, we need delicate fabrication techniques to set up a good junction. 
Another method to detect the node position is magnetic field orientation-sensitive 
experiments for physical quantities of bulk 
measurement.\cite{vekhter,miranovic2003,miranovic2005}  
There, by rotating magnetic field orientation, 
we try to identify the position of the nodes from the change of 
physical quantities, such as specific 
heat~\cite{t-park2003,aoki,deguchi,yamada,yano,t-park2008,sakakibara,an}, 
and thermal conductivity~\cite{izawa}, 
depending on the relative angle between 
magnetic field orientation and the node-direction.   
These quantities under magnetic fields reflect electronic excitation 
in the vortex states. 
For reliable analysis of the experimental data
and establishing those methods as a spectroscopic means of 
the gap node position determination, 
these electronic behaviors have to be confirmed by quantitative estimate 
of theoretical calculation in the vortex states.   
In this paper, we study the vortex states and 
the field-angle sensitive specific heat 
under magnetic fields parallel to the $ab$ plane, 
related to the pairing gap in a heavy fermion superconductor 
${\rm CeCoIn_5}$.  
For ${\rm CeCoIn_5}$, 
there were controversial discussions 
as for the pairing symmetry between $d_{x^2-y^2}$-wave and 
$d_{xy}$-wave~\cite{sakakibara}. 

\begin{figure}[tb]
\begin{center}

\includegraphics[width=5.0cm]{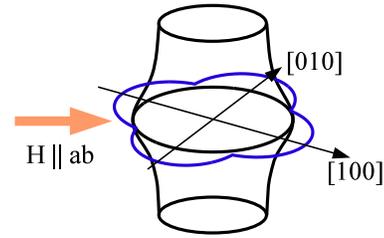}
\end{center}
  \caption{(Color online) 
Schematic configuration of magnetic field 
and $d_{x^2-y^2}$-wave superconducting gap 
on rippled cylindrical Fermi surface. 
We discuss the case when magnetic field is applied parallel to 
the $ab$ plane. 
}
\label{fig:FS}
\end{figure}

We consider the case when 
the magnetic field orientation is rotated within the $ab$-plane, 
and the Fermi-surface is quasi-two dimensional (Q2D), 
as shown in Fig. \ref{fig:FS}. 
In this $ab$ plane magnetic field configuration, 
we calculate the vortex structure selfconsistently 
with electronic states by quasiclassical Eilenberger theory. 
The selfconsistent calculation is necessary for quantitative estimate 
of physical quantities, because we have to determine the vortex core 
radius accurately.   
By these calculations, we investigate  
(1) the stable vortex lattice structure, 
(2) the local density of states (LDOS) of electrons in the vortex lattice state, 
and 
(3) the amplitude of specific heat oscillation under field rotation. 
We discuss their differences between magnetic field orientations 
${\bf H}\parallel{\rm node}$ and ${\bf H}\parallel{\rm antinode}$.  
In $d_{x^2-y^2}$-wave pairing, $[1, 1, 0]$ is node-direction, and 
$[1, 0, 0]$ is antinode-direction.

(1) 
The configuration of the vortex lattice reflects 
the anisotropy of superconductivity such as $d$-wave pairing. 
When ${\bf H}\parallel c$, 
due to the fourfold symmetry on the Fermi surface, 
triangular vortex lattice at low fields becomes the deformed lattice, 
accompanied by a first order transition of lattice orientation change, 
and finally reduces to a square vortex lattice~\cite{suzuki}.    
When ${\bf H}\parallel ab$, 
effects of anisotropic superconductivity on the vortex lattice configuration 
is not enough clarified. 
Thus, in this field direction, 
we discuss the transition between two possible vortex lattice configuration, 
depending $T$ or field orientation within the $ab$ plane. 

(2) 
In the LDOS when ${\bf H}\parallel c$, 
the zero-energy electronic states around vortex show 
star-shape spatial pattern, reflecting the anisotropy of the pairing gap function. 
The low-energy states extend from the vortex core toward the node 
(or minimum gap) directions.\cite{hess,hayashi} 
In this paper we discuss the LDOS around vortices when ${\bf H}\parallel ab$, 
both for ${\bf H}\parallel{\rm node}$ and ${\bf H}\parallel{\rm antinode}$.  
The LDOS structure mainly reflects properties of Q2D Fermi surface. 

(3) 
Since low temperature specific heat is proportional to 
zero-energy DOS $N(E=0)$, the oscillation of $N(E=0)$ 
by rotating magnetic field orientation was 
estimated by Doppler shift method~\cite{vekhter}  
and microscopical Eilenberger theory.~\cite{miranovic2003,miranovic2005} 
While the Doppler shift method \cite{volovik} is a handy estimate to 
intuitively understand the oscillation,
for quantitative comparison with experimental data,  
we have to carefully evaluate the oscillation amplitude by quantitatively reliable 
theoretical method such as Eilenberger theory,  
since the oscillation amplitude is in the order of a few percent
of the total specific heat value.  

In the early stage of the specific heat experiment for 
${\rm CeCoIn_5}$~\cite{aoki}, 
the sign of the oscillation was not consistent with the $d_{x^2-y^2}$-wave pairing. 
The $d_{x^2-y^2}$-wave pairing was suggested by 
some experiments, such as the oscillation of thermal conductivity by rotating 
magnetic fields,\cite{izawa}  
flux line lattice transformation,\cite{bianchi,kawamura,hiasa,suzuki} 
point-contact Andreev reflection,\cite{w-park} 
and spin resonance.\cite{stock,eremin} 
In order to settle the contradiction between the $d_{x^2-y^2}$-wave pairing 
and the specific heat oscillation, 
the possibility of the sign change as a function of $T$ 
was proposed.\cite{vorontsov2007,boyd} 
Since the sign of the oscillation is changed at intermediate $T$ range, 
the sign  can be opposite to that expected at low $T$. 
The previous work by Eilenberger theory for the sign change was done 
by the Pesch approximation,\cite{vorontsov2007} where the spatial dependence of the 
quasiclassical Green's function $g$ is neglected, 
replacing $g$ to the spatial average of $g$.   
In this paper, we quantitatively calculate the amplitude and sign 
of the specific heat oscillation by Eilenberger theory fully selfconsistently 
without using Pesch approximation. 
Extending this method further, we evaluate the contribution of strong paramagnetic 
effect and anisotropic Fermi surface velocity on the oscillation. 
The anisotropic Fermi velocity is another key factor other than 
the anisotropic superconducting gap, when we consider the 
vortex structure in anisotropic superconductors.       
The sign change of the specific heat oscillation at low $T$ 
and low $H$ has been recently observed in ${\rm CeCoIn_5}$, 
which confirms $d_{x^2-y^2}$-wave pairing.\cite{an} 
Part of the results in this paper was reported in ref. \citen{an} 
as theoretical analysis for the experiment.   
 
It is noted that the strong paramagnetic effect 
is important and indispensable when we discuss the vortex states in ${\rm CeCoIn_5}$.  
The paramagnetic effect comes from mismatched Fermi surfaces 
of up- and down-spin electrons due to the large Zeeman splitting.  
At higher fields in the case of strong paramagnetic effect,  
the upper critical field $H_{\rm c2}$ 
changes to the first order phase transition~\cite{izawa,bianch2002,tayama}  
and new Fulde-Ferrell-Larkin-Ovchinnikov (FFLO) state  may
appear.~\cite{ff,lo,bianchi2003,radovan,watanabe,capan,martin,kakuyanagi,kumagai,matsudaJPSJ}
Even in the vortex states at low field before entering to the FFLO state, 
the strong paramagnetic effects induces anomalous behavior of physical 
quantities~\cite{ikedaS,deguchi2007,tayama,ichiokapara}, including field dependence of 
flux line lattice form factor~\cite{DeBeer,bianchi}.  
Therefore, we consider contributions of the strong paramagnetic effect 
in the studies of items (1)-(3).

After giving our formulation of quasiclassical theory in the presence 
of the paramagnetic effect in \S \ref{sec:formulation}, 
we study the stable vortex lattice configuration
in \S \ref{sec:VL}  
and the zero-energy LDOS structure in \S \ref{sec:LDOS} 
for two field orientations 
${\bf H}\parallel{\rm node}$ and ${\bf H}\parallel{\rm antinode}$,  
when ${\bf H}\parallel ab$ in the $d_{x^2-y^2}$-wave pairing. 
In \S \ref{sec:C4}, we estimate the $H$- and $T$-dependences of 
amplitude and sign of specific heat oscillation by rotating  
the magnetic field orientation within the $ab$-plane, 
in the presence of strong paramagnetic effect and anisotropic 
Fermi velocity in addition to the $d$-wave pairing.   
The last section is devoted to summary and discussions.

\section{Formulation by Selfconsistent Quasiclassical Theory}
\label{sec:formulation}

We calculate the spatial structure of the vortex lattice state 
by quasiclassical Eilenberger theory in the clean 
limit.~\cite{eilenberger,kleinJLTP,ichiokaQCLs,ichiokaQCLd1}
The quasiclassical theory is quantitatively valid when 
$\xi \gg 1/k_{\rm F}$ ($k_{\rm F}$ is the Fermi wave number, and 
$\xi$ is the superconducting coherence length), 
which is satisfied in most of superconductors in solid states. 
When the paramagnetic effects are discussed, 
we include the Zeeman term $\mu_{\rm B}B({\bf r})$, 
where $B({\bf r})$ is the flux density of the internal field and 
$\mu_{\rm B}$ is a renormalized Bohr 
magneton.~\cite{WatanabeKita,klein,ichiokaFFLO,ichiokapara} 
The quasiclassical Green's functions
$g( \omega_n +{\rm i} {\mu} B, {\bf k},{\bf r})$, 
$f( \omega_n +{\rm i} {\mu} B, {\bf k},{\bf r})$, and 
$f^\dagger( \omega_n +{\rm i} {\mu} B, {\bf k},{\bf r})$  
are calculated in the vortex lattice state  
by the Eilenberger equation 
\begin{eqnarray} &&
\left\{ \omega_n +{\rm i}{\mu}B 
+\tilde{\bf v} \cdot\left(\nabla+{\rm i}{\bf A} \right)\right\} f
=\Delta\phi g, 
\nonumber 
\\ && 
\left\{ \omega_n +{\rm i}{\mu}B 
-\tilde{\bf v} \cdot\left( \nabla-{\rm i}{\bf A} \right)\right\} f^\dagger
=\Delta^\ast \phi^\ast g  , \quad 
\label{eq:Eil}
\end{eqnarray} 
where $g=(1-ff^\dagger)^{1/2}$, ${\rm Re} g > 0$, 
$\tilde{\bf v}={\bf v}/v_{{\rm F}0}$, 
and ${\mu}=\mu_{\rm B} B_0/\pi k_{\rm B}T_{\rm c}$. 
${\bf k}=(k_a,k_b,k_c)$  is the relative momentum of the Cooper pair, 
and ${\bf r}$ is the center-of-mass coordinate of the pair. 
We set the pairing function 
$\phi({\bf k})=\phi_{x^2-y^2}({\bf k})=
\sqrt{\mathstrut 2}(k_a^2-k_b^2)/(k_a^2+k_b^2)$ in $d_{x^2-y^2}$-wave pairing, 
and magnetic fields are applied to $[1,0,0]$ or $[1,1,0]$ directions 
in the crystal coordinate.    
For example, 
when a magnetic field is applied to $[1,0,0]$ direction,  
the coordinate $(x,y,z)$ for the vortex structure corresponds to 
$(b,c,a)$ of the crystal coordinate. 
In the case of $d_{xy}$-wave pairing  
$\phi({\bf k})=\phi_{xy}({\bf k})=2\sqrt{\mathstrut 2} k_a k_b /(k_a^2+k_b^2)$,   
the results for $[1,0,0]$ and $[1,1,0]$ field directions are exchanged. 

In our calculation, length, temperature, Fermi velocity, 
magnetic field and vector potential are, respectively, scaled by 
$R_0$, $T_c$, $\bar{v}_{\rm F}$, $B_0$ and $B_0 R_0$. 
Here, $R_0=\hbar \bar{v}_{\rm F}/2 \pi k_{\rm B} T_{\rm c}$, 
$B_0=\phi_0 /2 \pi R_0^2$ with the flux quantum $\phi_0$, 
and $\bar{v}_{\rm F}=\langle v_{\rm F}^2 \rangle_{\bf k}^{1/2}$ 
is an averaged Fermi velocity on the Fermi surface. 
$\langle \cdots \rangle_{\bf k}$ indicates the Fermi surface average. 
The energy $E$, pair potential $\Delta$ and Matsubara frequency $\omega_l$ 
are in unit of $\pi k_{\rm B} T_{\rm c}$. 
As a model of the Fermi surface, for simplicity, 
we use a Q2D Fermi surface with rippled cylinder-shape, 
and the Fermi velocity is given by 
\begin{eqnarray}
{\bf v}_{\rm F}=(v_a,v_b,v_c) \propto 
(\tilde{v} \cos\theta_k, \tilde{v}\sin\theta_k, \tilde{v}_z \sin k_c)
\label{eq:vf}
\end{eqnarray}
at the Fermi surface 
${\bf k}_{\rm F}=(k_a,k_b,k_c)
\propto(k_{\rm F0}\cos\theta_k,k_{\rm F0}\sin\theta_k,k_c)$ 
with $-\pi \le \theta_k \le \pi$ and $-\pi \le k_c \le \pi$.~\cite{ichiokaMgB2} 
To include the Fermi velocity anisotropy, we use 
$\tilde{v}=1+\beta \cos 4 \theta_k$. 
In our calculation except for last part, $\beta=0$ 
since we mainly consider the case when the Fermi velocity is isotropic in the $ab$-plane. 
In our work,  we set $\tilde{v}_z =0.5$, so that the anisotropy ratio 
\begin{eqnarray}
\gamma=\frac{\xi_c}{\xi_{ab}} \sim  
\frac{\langle v_c^2 \rangle_{\bf k}^{1/2}}{\langle v_a^2 \rangle_{\bf k}^{1/2}} 
\sim 0.5, 
\label{eq:gamma}
\end{eqnarray} 
as in ${\rm CeCoIn_5}$.

When magnetic fields are applied to the $z$ axis of the vortex coordinate,  
the vector potential 
${\bf A}({\bf r})=\frac{1}{2} {\bf H} \times {\bf r}
 + {\bf a}({\bf r})$ in the symmetric gauge, 
where ${\bf H}=(0,0,H)$ is a uniform flux density and 
${\bf a}({\bf r})$ is related to the internal field 
${\bf B}({\bf r})={\bf H}+\nabla\times {\bf a}({\bf r})$.
The unit cell of the vortex lattice is given by 
${\bf r}=s_1({\bf u}_1-{\bf u}_2)+s_2{\bf u}_2$ with 
$-0.5 \le s_i \le 0.5$ ($i$=1, 2), ${\bf u}_1=(a_x,0,0)$,   
${\bf u}_2=(a_x/2,a_y,0)$ and $a_x a_y B=\phi_0$.  

As for selfconsistent conditions, 
the pair potential is calculated by 
\begin{eqnarray}
\Delta({\bf r})
= g_0N_0 T \sum_{0 < \omega_n \le \omega_{\rm cut}} 
 \left\langle \phi^\ast({\bf k}) \left( 
    f +{f^\dagger}^\ast \right) \right\rangle_{\bf k} 
\label{eq:scD} 
\end{eqnarray} 
with 
$(g_0N_0)^{-1}=  \ln T +2 T
        \sum_{0 < \omega_n \le \omega_{\rm cut}}\omega_l^{-1} $. 
We use $\omega_{\rm cut}=20 k_{\rm B}T_{\rm c}$.
The vector potential for the internal magnetic field 
is selfconsistently determined by 
\begin{eqnarray}
\nabla\times \left( \nabla \times {\bf A} \right) 
=\nabla\times {\bf M}_{\rm para}({\bf r})
-\frac{2T}{{{\kappa}}^2}  \sum_{0 < \omega_n} 
 \left\langle {\bf v}_{\rm F} 
         {\rm Im} g  
 \right\rangle_{\bf k}, 
\label{eq:scH} 
\end{eqnarray} 
where we consider both the diamagnetic contribution of 
supercurrent in the last term 
and the contribution of the paramagnetic moment 
${\bf M}_{\rm para}({\bf r})=(0,0,M_{\rm para}({\bf r}))$ 
with 
\begin{eqnarray}
M_{\rm para}({\bf r})
=M_0 \left( 
\frac{B({\bf r})}{H} 
- \frac{2T}{{\mu} H }  
\sum_{0 < \omega_n}  \left\langle {\rm Im} \left\{ g \right\} 
 \right\rangle_{\bf k}
\right) . 
\label{eq:scM} 
\end{eqnarray} 
The normal state paramagnetic moment 
$M_0 = ({{\mu}}/{{\kappa}})^2 H $,   
${\kappa}=B_0/\pi k_{\rm B}T_{\rm c}\sqrt{8\pi N_0}$  and 
$N_0$ is the DOS at the Fermi energy in the normal state. 
We set the Ginzburg-Landau parameter $\kappa=89$ for this 
typical type-II superconductor.  
We solve eq. (\ref{eq:Eil}) and eqs. (\ref{eq:scD})-(\ref{eq:scM})
alternately, and obtain selfconsistent solutions
as in previous works,~\cite{ichiokaQCLs,ichiokaQCLd1,ichiokapara}
under a given unit cell of the vortex lattice. 

In Eilenberger theory, free energy is given by 
\begin{eqnarray} &&  
F=\int_{\rm unit cell} {\rm d}{\bf r} \Bigl\{ 
{\kappa}^2 |{\bf B}({\bf r})-{\bf H}|^2 -{\mu}^2 |B({\bf r})|^2 
\nonumber \\ && \qquad
+ |\Delta({\bf r})|^2 ( 
  \ln T + 2 T \sum_{0<\omega_n<\omega_{\rm cut}} \omega_n^{-1} )
\nonumber \\ && \qquad
-T  \sum_{|\omega_n|<\omega_{\rm cut}}
   \left\langle I({\bf r},{\bf k},\omega_n) \right\rangle 
\Bigr\}  
\label{eq:f1}
\end{eqnarray}
with 
\begin{eqnarray} &&  
I({\bf r},{\bf k},\omega_n)
=\Delta \phi f^\dagger + \Delta^\ast \phi^\ast f 
\nonumber \\ && \qquad
+(g -\frac{\omega_n}{|\omega_n|})
\Bigl\{ \frac{1}{f}\left( \omega_n +{\rm i}{\mu}B
      + \hat{\bf v}\cdot(\nabla+{\rm i}{\bf A}) \right)f
\nonumber \\ && \qquad\qquad
+ \frac{1}{f^\dagger}\left( \omega_n +{\rm i}{\mu}B
      + \hat{\bf v}\cdot(\nabla-{\rm i}{\bf A}) \right)f^\dagger 
\Bigr\} 
\label{eq:f2}
\end{eqnarray}
in our dimensionless unit.
Using eqs. (\ref{eq:Eil}) and (\ref{eq:scD}), we obtain 
\begin{eqnarray} &&
F=
\int_{\rm unit cell} {\rm d}{\bf r} \Bigl\{ 
{\kappa}^2 |{\bf B}({\bf r})-{\bf H}|^2 -{\mu}^2 |B({\bf r})|^2 
\nonumber \\ && \qquad
+T  \sum_{|\omega_n|<\omega_{\rm cut}}
  {\rm Re} \left\langle 
\frac{g-1}{g+1}(\Delta \phi f^\dagger + \Delta^\ast \phi^\ast f )
 \right\rangle 
\Bigr\}  .
\label{eq:f3}
\end{eqnarray} 
The entropy in the superconducting state,  
given by $S_{\rm s}(T)=S_{\rm n}(T)- \partial F/\partial T$,  
is obtained from eqs. (\ref{eq:f1}) and (\ref{eq:f2}) as  
\begin{eqnarray} && 
\frac{S_{\rm s}(T)}{S_{\rm n}(T_{\rm c})}
=T - \frac{3}{2} \int_{\rm unit cell} {\rm d}{\bf r} \Bigl\{ 
\nonumber \\ && 
( \ln T + 2 T \sum_{0<\omega_n<\omega_{\rm cut}} \omega_n^{-1} )^{-1}
\sum_{0<\omega_n<\omega_{\rm cut}}
{\rm Re}\langle \Delta \phi f^\dagger +\Delta^\ast \phi^\ast f \rangle 
\nonumber \\ && 
-2 \sum_{0<\omega_n<\omega_{\rm cut}}( {\rm Re} \left\langle 
\frac{ \Delta \phi f^\dagger +\Delta^\ast \phi^\ast f}{g+1} \right\rangle 
+2 \omega_n {\rm Re}\langle g-1 \rangle ) 
\Bigr\} 
\qquad
\label{eq:S}
\end{eqnarray}
in our dimensionless unit. 
$S_n$ is the entropy in the normal state.
We calculate $S_{\rm s}(T)$ numerically using selfconsistent solutions of 
quasiclassical Green's functions, $\Delta({\bf r})$ and ${\bf A}({\bf r})$. 
By numerical derivative of $S_{\rm s}(T)$, we obtain the specific heat $C$ as 
\begin{eqnarray}
C=T \frac{\partial S_{\rm s}} {\partial T}. 
\label{eq:C}
\end{eqnarray}

\section{Stable Vortex Lattice Configuration When $H \parallel ab$}
\label{sec:VL}

\begin{figure}[tb]
\begin{center}
\begin{minipage}{4.5cm}
\includegraphics[width=4.5cm]{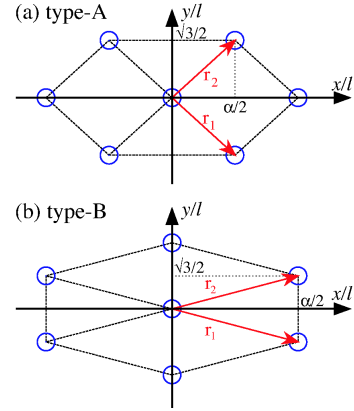}
\end{minipage}
\vspace{5mm}\\
\begin{minipage}{8cm}
\includegraphics[width=8.0cm]{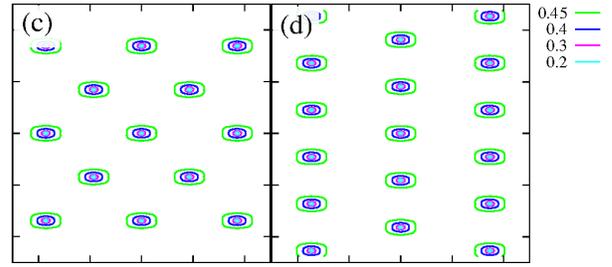}
\end{minipage}

\end{center}
  \caption{(Color online) 
Possible stable vortex lattice configurations type-A (a) and type-B (b) 
when ${\bf H}\parallel ab$.  
(c) and (d) show
the spatial structure of the pair potential $|\Delta({\bf r})|$ 
in two vortex lattice configurations $\alpha=1.9$ (a) and 6.6 (b) 
for ${\bf H}\parallel {\rm antinode}$.  
$\mu=2$, $H/H_{\rm c2}=0.211$ and $T/T_{\rm c}=0.2$. 
Both (c) and (d) panels, the horizontal axis is [100] direction, and 
the vertical axis is [001] direction, and the view range is 25$\times$25
in the Eilenberger  length unit $R_0$.
}
\label{fig:latticeD}
\end{figure}

When ${\bf H} \parallel ab$, in the absence of paramagnetic effect ($\mu=0$), 
the upper critical field $H_{\rm c2}$ has field orientation dependence;  
$H_{\rm c2}\sim 1.55$ for ${\bf H} \parallel {\rm antinode}$ 
and $H_{\rm c2}\sim 1.45$ for ${\bf H} \parallel {\rm node}$ 
in our parameters. 
In the case of strong paramagnetic effect $\mu=2$, 
$H_{\rm c2}$ is largely suppressed to $H_{\rm c2}\sim 0.190$ 
both for ${\bf H} \parallel {\rm antinode}$ and ${\bf H} \parallel {\rm node}$. 
There orientation-dependence of $H_{\rm c2}$ becomes isotropic within the $ab$ plane.   

First, we discuss the vortex lattice configuration,  
when magnetic fields are applied parallel to the $ab$ plane, i.e., 
${\bf H}\parallel ab$. 
In these field orientations, 
there are two candidates for the stable vortex lattice configuration. 
One is when one of the neighbor six vortices is located in the $ab$-direction, 
as shown in Fig. \ref{fig:latticeD} (a), 
which is called type-A configuration in this paper.  
In the other configuration, 
one of the neighbor vortices is located in the $c$ direction,    
as shown in Fig. \ref{fig:latticeD} (b), called type-B configuration.  
In the isotropic case of the $s$-wave pairing and the Fermi sphere, 
anisotropy ratio of the vortex lattice defined by $\alpha=2 a_y /\sqrt{3}a_x$ 
is $\alpha=1$ in type-A, and $\alpha=3$ in type-B. 
In our parameter where anisotropic ratio of Fermi velocity, $\gamma^{-1}\sim 2$, 
the triangular lattice is distorted by this ratio so that 
$\alpha$ for stable vortex lattice becomes about two times larger, i.e., 
$\alpha \sim 2$ or 6. 
The exact value of $\alpha$ is not trivial because of the contributions of rippled 
cylindrical Fermi surface 
and line nodes of $d$-wave pairing. 
Thus, we have to estimate $\alpha$ for stable vortex lattice configuration.   

\begin{figure}[tb]
\begin{center}
\includegraphics[width=6.0cm]{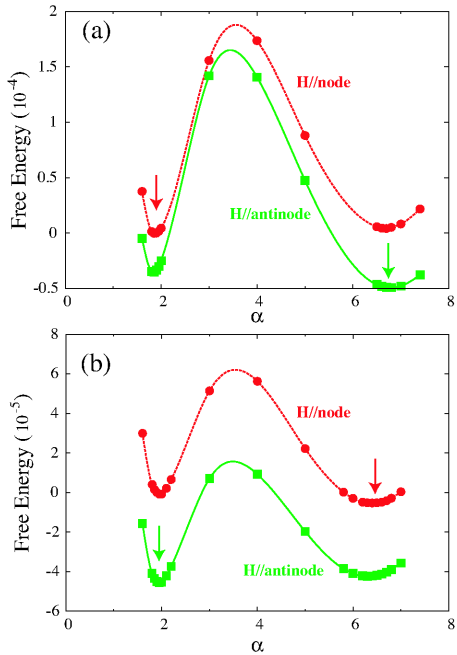}
\end{center}
\caption{(Color online) 
Free energy $F-F_0$ as a function of $\alpha$ 
at $T/T_{\rm c}=0.1$ (a) and 0.6 (b) 
in the presence of strong paramagnetic effect ($\mu=2$). 
$H/H_{\rm c2}=0.211$. 
$F_0 \equiv F(\alpha=1.9,H \parallel {\rm node}) $.  
In each figure, the upper (lower) line shows the case when 
magnetic field is applied to the node (antinode) direction of 
the $d$-wave pairing gap function.  
For each field orientation, 
the arrow indicates the minimum of $F$. 
}
\label{fig:F-alpha}
\end{figure}

In Fig. \ref{fig:F-alpha}, we plot the free energy $F$ 
as a function of $\alpha$ 
for two magnetic field orientations  
at $T/T_{\rm c}=0.1$ and 0.6 in the presence of large paramagnetic effect 
($\mu=2$). 
There, the local minimum for $F$ is located at two configurations 
$\alpha \sim 1.9$ (type-A) and 6.6 (type-B). 
Among two local minima, vortex lattice configuration with lower $F$ is stable, 
and the other is metastable. 
When the field is applied to the node direction of the $d$-wave pairing function, 
stable vortex lattice is type-A at $T/T_{\rm c}=0.1$, and changed to type-B 
at $T/T_{\rm c}=0.6$. 
On the other hand, when the field is parallel to antinode direction,   
stable vortex lattice is type-B at $T/T_{\rm c}=0.1$, and changed to type-A 
at $T/T_{\rm c}=0.6$. 
Therefore, the vortex lattice configuration can be changed between 
type-A and type-B, as a first order transition by rotation magnetic 
field orientation within the $ab$ plane.  

\begin{figure}[tb]
\begin{center}
\includegraphics[width=6.0cm]{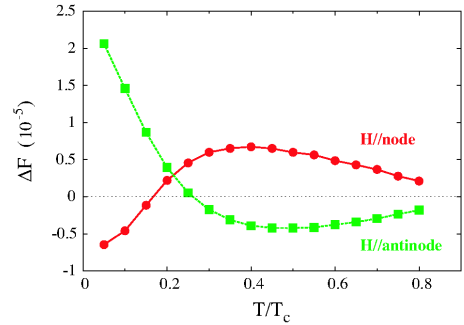}
\end{center}
\caption{(Color online) 
Temperature-dependence of the free energy difference $\Delta F$
between two vortex lattice configurations type-A and type-B. 
We plot $\Delta F \equiv F(\alpha=1.9)-F(\alpha=6.6)$,  
as a function of $T$,  
for two magnetic field orientations 
${\bf H} \parallel {\rm antinode}$ (squares) and ${\bf H} \parallel {\rm node}$ (circles). 
$\mu=2$ and $H/H_{\rm c2}=0.211$.  
}
\label{fig:T-dF}
\end{figure}

To see the temperature dependence, 
in Fig. \ref{fig:T-dF}
we show the free energy difference 
$\Delta F$ between vortex lattice configurations type-A ($\alpha=1.9$) 
and type-B ($\alpha=6.6$). 
When $\Delta F>0$, type-B is stable, 
and when $\Delta F<0$, type-A is stable.  
From Fig. \ref{fig:T-dF}, we see that 
stable vortex lattice configuration is changed at $T \sim 0.2 T_{\rm c}$. 
Near $T \sim 0.2 T_{\rm c}$
vortex lattice configuration of type-B is stable both 
field orientations ${\bf H} \parallel {\rm node}$ and  
${\bf H} \parallel {\rm antinode}$. 
As seen in Fig. \ref{fig:F-alpha}, 
at lower (higher) temperature regions  
stable vortex lattice configuration is type-A (type-B) 
for ${\bf H} \parallel {\rm node}$, while  
the configuration of type-B (type-A) is stable 
for ${\bf H} \parallel {\rm antinode}$.  
Thus, the transition of vortex lattice between type-A and type-B 
can occur upon changing temperature. 

\begin{figure}[tb]
\begin{center}
\includegraphics[width=6.0cm]{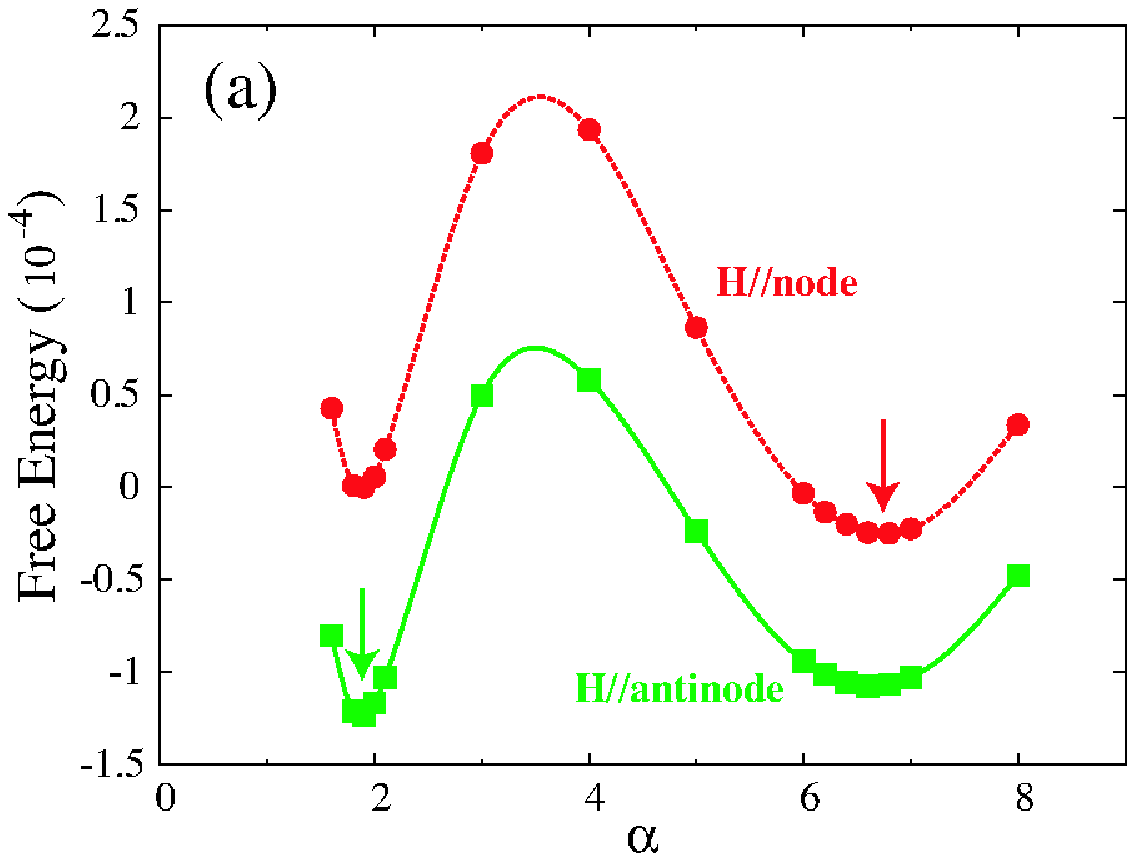}
\includegraphics[width=6.0cm]{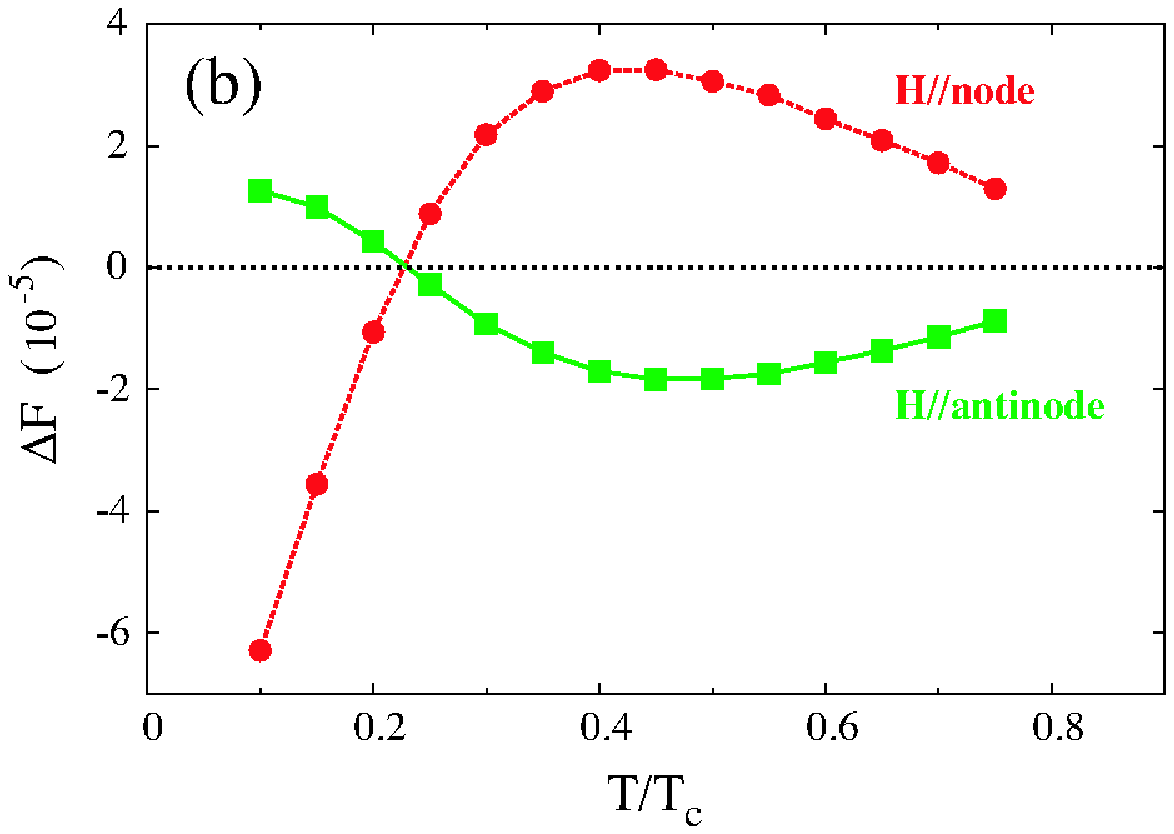}
\end{center}
\caption{(Color online) 
(a) 
Free energy $F-F_0$ as a function of $\alpha$ 
at $T/T_{\rm c}=0.6$. 
$F_0 \equiv F(\alpha=1.9,H \parallel {\rm node}) $.  
For each field orientation, 
the arrow  indicates the minimum of $F$. 
(b)  
$T$-dependence of the free energy difference 
$\Delta F \equiv F(\alpha=1.9)-F(\alpha=6.6)$, 
between two vortex lattice configurations type-A and type-B. 
Here we show the cases without paramagnetic effect ($\mu=0$) 
for magnetic field orientations ${\bf H} \parallel {\rm antinode}$ (squares)
and ${\bf H} \parallel {\rm node}$ (circles). 
$H/H_{\rm c2}\sim 0.065$ with $H_{\rm c2}\equiv H_{\rm c2,antinode}\sim 1.55$. 
}
\label{fig:T-dFmu0}
\end{figure}

For reference, in Fig. \ref{fig:T-dFmu0}, 
we show the results when the paramagnetic effect is absent 
($\mu=0$). 
There, the results are qualitatively unchanged from those of 
Figs. \ref{fig:F-alpha} and \ref{fig:T-dF} in the strong paramagnetic case.  
The free energy as a function of $\alpha$ has local minimum 
at $\alpha \sim 1.9$ and $\alpha \sim 6.6$. 
At lower $T$, the stable vortex lattice configuration is 
type-A for ${\bf H} \parallel {\rm node}$, 
and 
type-B for ${\bf H} \parallel {\rm antinode}$. 
These are changed at higher $T$.  

The purpose of this section was that we proposed the possibility of 
interesting transitions in the  stable vortex lattice configuration 
when magnetic field is applied to the $ab$ plane. 
The vortex lattice configuration can be changed between 
type-A and type-B, depending on $T$, $H$, and 
the field orientation relative to the node direction of 
the $d$-wave pairing.~\cite{white} 
It is noted that 
the results of the stable vortex lattice shown in our calculation 
is an example for the transition of the vortex lattice configuration.  
It is not sure that these results of $T$-dependence are universal.  
Since the free energy difference $\Delta F$ is very small, 
it is difficult to identify the reason for changes of 
stable vortex lattice configurations.  
Detailed forms of Fermi surface shape and anisotropic pairing gap function may 
change the results for the stable vortex lattice configuration.  
Therefore, 
we have to carefully estimate it, using realistic Fermi surface, 
to compare with future experimental data.  
However, it is interesting to experimentally 
identify the phase diagram of the stable vortex lattice configuration 
under parallel field to the $ab$ plane.

\section{Electronic States in Vortex Lattice When ${\bf H} \parallel ab$}
\label{sec:LDOS}

\begin{figure}[tb]
\begin{center}
\includegraphics[width=9.0cm]{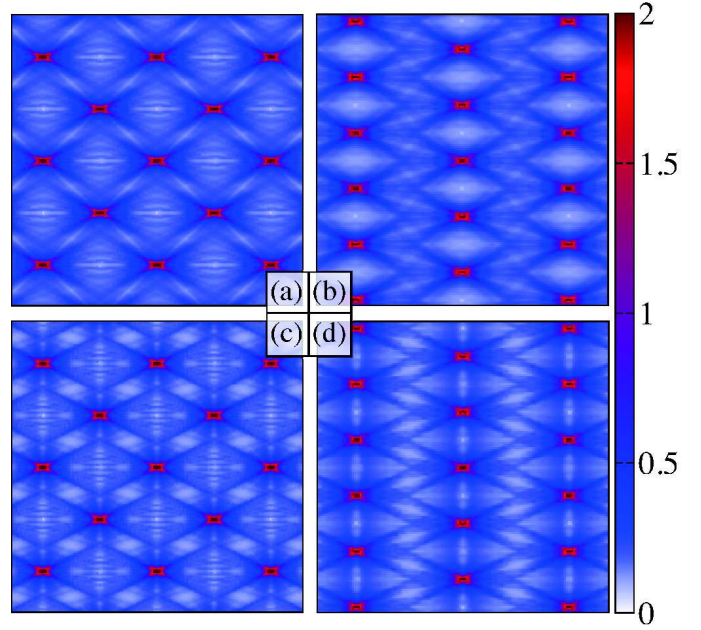}
\end{center}
\caption{(Color online) 
Spatial structure of zero-energy LDOS $N(E=0,{\bf r})$ 
for $\mu=0$ at $H/H_{\rm c2}=0.065$
in the vortex lattice under magnetic fields parallel to the $ab$ plane. 
Upper two panels (a) and (b) are for a magnetic field 
along the antinode direction of the $d$-wave paring gap function.
The vertical axes are [001] directions and the horizontal axes are [100] directions.
Lower two panels (c) and (d) are for the node-direction. 
The vertical axes are [001] directions and the horizontal axes are [110] directions.
Left-side two panels (a) and (c) are for the vortex lattice configuration 
of type-A ($\alpha=1.9$). 
Right-side two panels (b) and (d) are for type-B configuration ($\alpha=6.6$). 
The view ranges of all four panels are 15$\times$15 in the Eilenberger length unit $R_0$.

}
\label{fig:E0LDOSmu0}
\end{figure}

\begin{figure}[tb]
\begin{center}
\includegraphics[width=9.0cm]{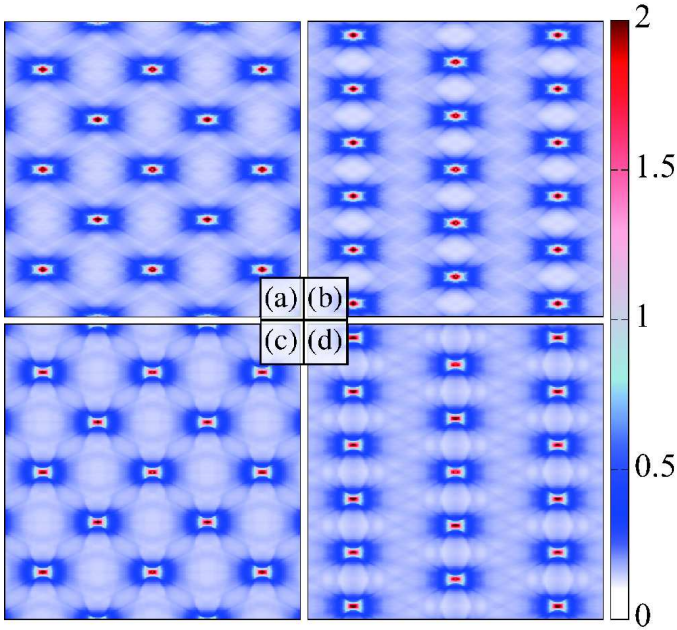}
\end{center}
\caption{(Color online) 
The same as in Fig. \ref{fig:E0LDOSmu0}, but in the presence of 
strong paramagnetic effect ($\mu=2$) and $H/H_{\rm c2}=0.211$.  
The view ranges of all four panels are 25$\times$25 in the Eilenberger length unit $R_0$.
}
\label{fig:E0LDOSmu2}
\end{figure}

When we calculate the electronic states,
we solve eq. (\ref{eq:Eil}) with 
$ {\rm i}\omega_n \rightarrow E + {\rm i} \eta$.
The LDOS is given by
$N({\bf r},E)=N_{\uparrow}({\bf r},E)+N_{\downarrow}({\bf r},E)$, where 
\begin{eqnarray}
N_\sigma({\bf r},E)=N_0 \langle {\rm Re }
\{
g( \omega_n +{\rm i} \sigma{\mu} B, {\bf k},{\bf r})
|_{i\omega_n \rightarrow E + i \eta} \}\rangle_{\bf k}
\end{eqnarray}
with $\sigma=1$ ($-1$) for up (down) spin component. 
We typically use $\eta=0.01$.
The DOS is obtained by the spatial average of the LDOS as 
$N(E)=N_\uparrow (E) +N_\downarrow (E)
 =\langle N({\bf r},E) \rangle_{\bf r}$. 

To show the electronic states in the vortex lattice for ${\bf H}\parallel ab$ 
without the paramagnetic effect,  
in Fig. \ref{fig:E0LDOSmu0} we present zero-energy LDOS $N({\bf r},E=0)$ for 
two vortex lattice configurations type-A and type-B for two magnetic field 
orientations ${\bf H}\parallel{\rm node}$ and  ${\bf H}\parallel{\rm antinode}$.
There, zero-energy electronic states localized around vortex cores have tails 
extending outside the core. 
In our calculation, since we assume an open Fermi surface of rippled cylinder, 
there are not Fermi velocity pointing to the $c$-axis directions. 
Therefore quasiparticles with the $ab$-direction component Fermi velocity dominantly  
contribute to the zero-energy LDOS. 
Therefore,  $N({\bf r},E=0)$ extends to $ab$ direction from vortex cores, 
reflecting the rippled open Fermi surface.  
And the connection of zero-energy electronic states between vortices 
are week along the $c$-axis direction, compared to other directions. 
Two parallel lines of the LDOS tails connecting neighbor vortices are 
due to the interference between electrons at neighbor vortex cores.  
When zero-energy LDOS are connected between the vortices,  
the LDOS is suppressed just on the straight line directly 
connecting the vortex centers, 
resulting in two parallel tails of the zero-energy LDOS.~\cite{ichiokaQCLs}  
As for the difference between 
${\bf H}\parallel{\rm node}$ and  ${\bf H}\parallel{\rm antinode}$, 
zero-energy states around vortex core are broadly extended towards outside 
when ${\bf H}\parallel{\rm node}$,  
because superconducting gap has node when quasi-particles propagate to the direction 
perpendicular to the magnetic field orientation. 
Zero energy LDOS in the presence of strong paramagnetic effect ($\mu=2$) is 
presented in Fig. \ref{fig:E0LDOSmu2}. 
Due to the Zeeman splitting by strong paramagnetic effect, 
the original zero-energy bound state for $\mu=0$ is lifted 
to finite energy.\cite{ichiokapara} 
Thus, the tails extending from vortex cores are smeared in
 zero-energy LDOS of Fig. \ref{fig:E0LDOSmu2}.

\section{Field-Angle Dependence of Specific Heat}
\label{sec:C4}

\begin{figure}[tb]
\begin{center}
\includegraphics[width=7.0cm]{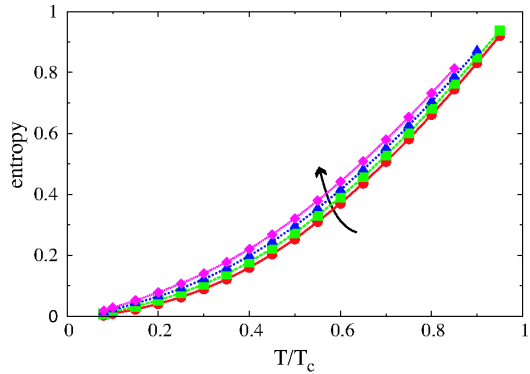}
\end{center}
\caption{(Color online) 
$T$-dependence of entropy $S_s(T)/S_n(T_c)$ at $H/H_{c2}=$0.105, 0.211, 0.316, and 0.421
from bottom to top.
$\mu=2$. 
}
\label{fig:T-Smu2}
\end{figure}

\begin{figure}[tb]
\begin{center}
\includegraphics[width=8cm]{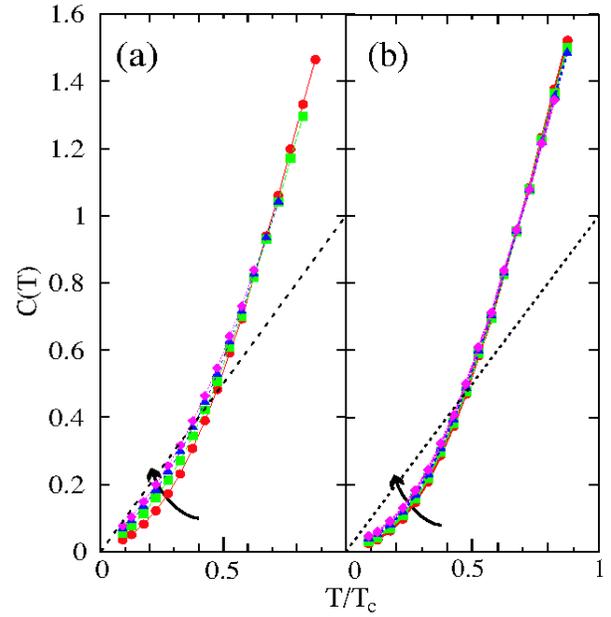}
\end{center}
\caption{(Color online) 
$T$-dependence of Specific heat $C$ 
at $H/H_{\rm c2}=$0.065, 0.194, 0.323, and 0.452 from bottom to top
for $\mu=0$ (a), 
and 
at $H/H_{\rm c2}=$0.105, 0.211, 0.316, and 0.421 from bottom to top for
$\mu=2$ (b).
Dotted line indicates $C(T)$ in the normal state. 
}
\label{fig:T-Cmu2}
\end{figure}

Using selfconsistent results of $\Delta({\bf r})$, 
${\bf A}({\bf r})$ and quasiclassical Green's functions, 
we calculate the entropy $S_{\rm s}(T)$ by eq. (\ref{eq:S}).
In Fig. \ref{fig:T-Smu2}, we show $T$-dependence of $S_{\rm s}(T)$. 
From the $T$-dependence, we numerically obtain $T$-dependence of 
the specific heat by eq. (\ref{eq:C}), which is presented in 
Fig. \ref{fig:T-Cmu2}. 
There, $C \propto T^2 $ at low fields because of line nodes in the $d$-wave pairing.
As shown in  Fig. \ref{fig:T-Cmu2}(a) when paramagnetic effect is negligible 
($\mu=0$), 
with increasing $H$, $C$ reduces to $T$-linear behavior 
due to low energy excitations around vortices,  
and approaches the line for the normal states.   
Since $T_{\rm c}$ decreases with increasing $H$, 
the jump of $C$ at $T_{\rm c}$ becomes smaller at high fields. 
When paramagnetic effect is strong ($\mu=2$) as shown 
in Fig. \ref{fig:T-Cmu2}(b), 
$C$ does not deviate from the low field curve 
even at $H/H_{\rm c2}=0.42$,  
because $H(=0.08)$ is still small. 
It changes to the normal state by the first order transition at $H_{\rm c2}$.

\begin{figure}[tb]
\begin{center}
\includegraphics[width=7.0cm]{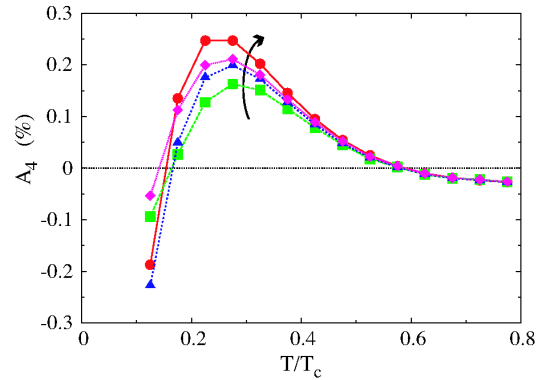}
\end{center}
\caption{(Color online)
$T$-dependence of fourfold oscillation part $A_4$ in specific heat 
under magnetic field rotation 
at $\mu=2$ and $H/H_{\rm c2}=0.211$. 
We show $A_4$ for some vortex lattice configurations 
$(\alpha_{\rm node},\alpha_{\rm antinode})=$(6.6,6.6),  (1.9,6.6), (6.6,1.9), (1.9,1.9)
from bottom to top.
}
\label{fig:T-C4mu2alpha}
\end{figure}
\begin{figure}[tb]
\begin{center}
\includegraphics[width=6.0cm]{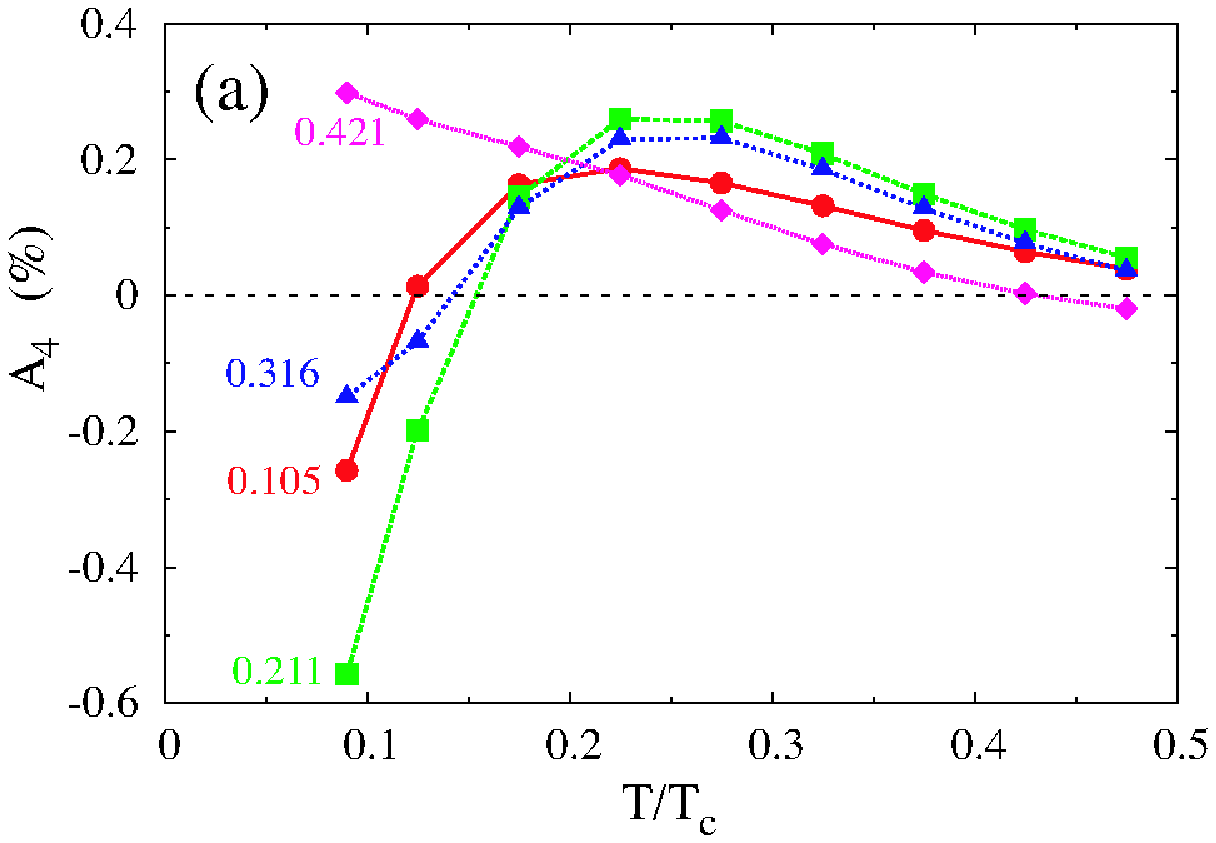}
\includegraphics[width=6.0cm]{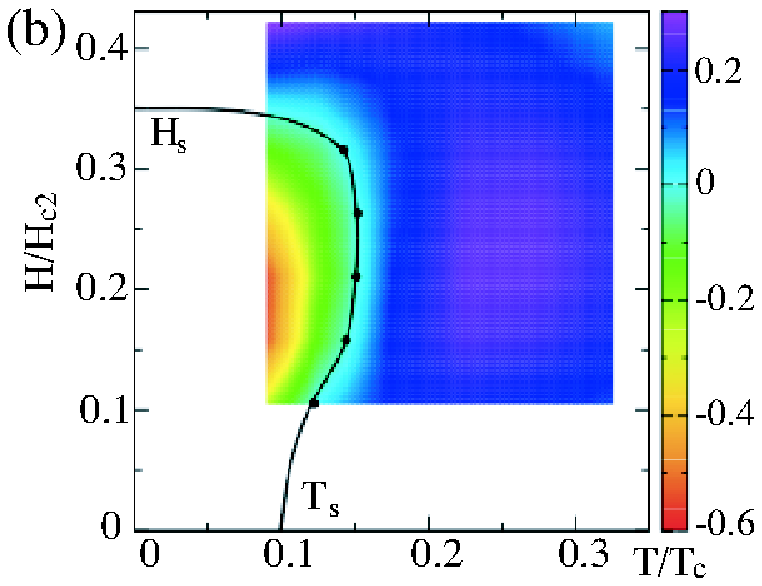}
\end{center}
\caption{(Color online) 
Fourfold oscillation part $A_4$ of specific heat 
in the case of strong paramagnetic effect $\mu=2$.
(a) $T$-dependence of oscillation part $A_4$ 
at $H/H_{\rm c2}=$0.105, 0.211, 0.316, and 0.421.
(b) $(T,H)$-dependence of $A_4$.  
The solid curve indicate position of $A_4=0$ where sign of the oscillation changes. 
$H_{\rm c2}\sim 0.19$. 
$\beta=0$. 
}
\label{fig:T-C4mu2}
\end{figure}

Next, we discuss the oscillation of the specific heat 
in the form 
\begin{eqnarray}
C=C_{H}(1-A_4 \cos 4 \theta ), 
\label{eq:A4}
\end{eqnarray}
when magnetic field orientation is rotated within $ab$-plane. 
$\theta$ is a relative angle between field orientation and antinode-direction 
of the $d$-wave gap function. 
The oscillation factor $A_4$ is calculated by 
\begin{eqnarray}
A_4=\frac{C_{\rm node}-C_{\rm antinode}}{C_{\rm node}+C_{\rm antinode}}
\times 100 \ [{\%}],  
\end{eqnarray}
where $C_{\rm node}$ ($C_{\rm antinode}$) is the specific heat 
for field orientation along the node (antinode) direction, 
i.e., $\theta=45^\circ$ ($0^\circ$) in the $d_{x^2-y^2}$-wave pairing. 

To evaluate how $A_4$ depends on the vortex lattice configurations 
type-A and type-B, 
in Fig. \ref{fig:T-C4mu2alpha} we present $T$-dependence of $A_4$ 
for some choices $(\alpha_{\rm node},\alpha_{\rm antinode})=$
(1.9,1.9), (6.6,6.6), (1.9,6.6), (6.6,1.9). 
$\alpha_{\rm node}$ ($\alpha_{\rm antinode}$) is value of $\alpha$ 
for vortex lattice configuration when field orientation is along the node 
(antinode) direction. 
From the figure, we see that the choice of the stable vortex lattice configuration 
 does not seriously affect on the behavior of $A_4$. 
In any cases, on lowering $T$ from high temperature, 
$A_4$ increases with sign change from negative to positive, 
and after peak near $T \sim 0.3 T_{\rm c}$, 
$A_4$ rapidly decreases with sign change to negative. 
Therefore, we fix the vortex lattice configuration as 
type-A with $\alpha=2$, hereafter. 
We note that $C_4$ for $\alpha=2$ does not seriously change 
from that for $\alpha=1.9$.   

Figure \ref{fig:T-C4mu2}(a) shows $T$-dependence of $A_4$ for some $H$. 
There, for lower $H$, sign change of $A_4$ occurs at $T \sim 0.15 T_{\rm c}$.    
However, at higher $H$, $A_4$ remains positive and increases on lowering $T$. 
To show the $H$- and $T$-dependence, 
we show contour plot of $A_4(T,H)$ in Fig. \ref{fig:T-C4mu2}(b).
There solid line indicate the sign change of $A_4(T,H)$. 
Inside region from the solid line at lower $T$ and lower $H$ has negative $A_4$. 
Other higher $T$ region has positive $A_4$. 
We defined the magnetic field where the sign change occurs at low temperature 
as $H_{\rm s}$, 
and the temperature of the sign change as $T_{\rm s}$. 
In Fig. \ref{fig:T-C4mu2}(b) in the presence of strong paramagnetic effect, 
$T_{\rm s}/T_{\rm c} \sim 0.1$ and $H_{\rm s}/H_{\rm c2} \sim 0.35$ 
[$H_{\rm s}\sim 0.067$]. 

\begin{figure}[tb]
\begin{center}
\includegraphics[width=6.0cm]{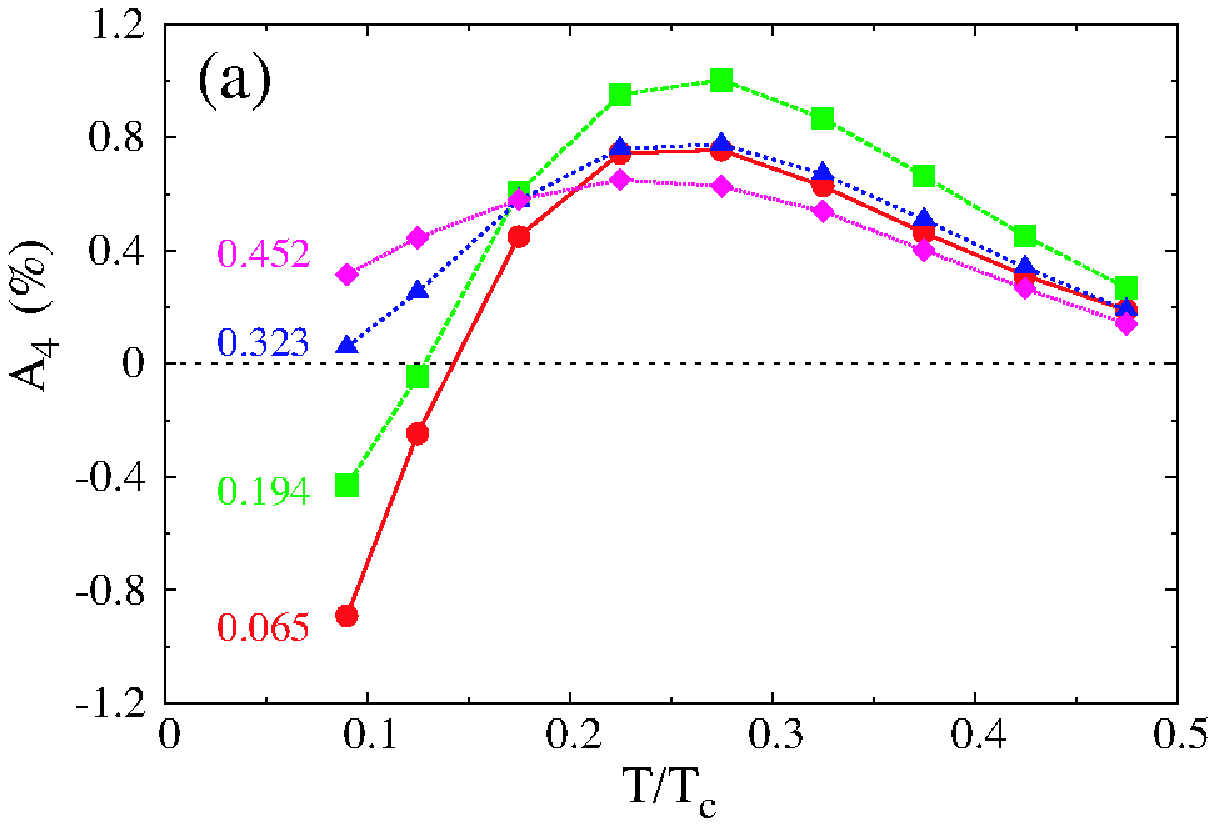}
\includegraphics[width=6.0cm]{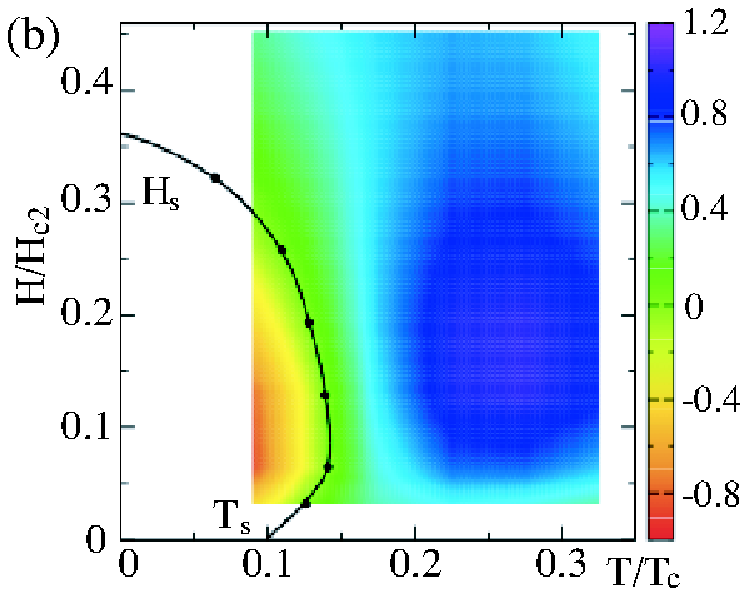}
\end{center}
\caption{(Color online) 
The same as Fig. \ref{fig:T-C4mu2}, 
but in the case when the paramagnetic effect is absent ($\mu=0$). 
In (a), $H/H_{\rm c2}=$0.065,  0.194, 0.323, and 0.452.
$H_{\rm c2} \equiv H_{\rm c2,antinode}\sim 1.55$. 
$\beta=0$. 
}
\label{fig:T-C4mu0}
\end{figure}

To evaluate the contribution of Pauli-paramagnetic effect, 
in Fig. \ref{fig:T-C4mu0} 
we show $A_4(T,H)$ in the case without paramagnetic effect ($\mu=0$). 
There, sign change of $A_4$ similarly occurs at low $T$ and $H$, 
while amplitude of $A_4$ is enhanced compared with that for $\mu=2$ 
in Fig. \ref{fig:T-C4mu2}. 
This case ($\mu=0$) corresponds to the previous work in ref. \citen{vorontsov2007}. 
Our results are qualitatively consistent to it, 
while we perform selfconsistent calculation without using Pesch approximation. 
Quantitatively, $H_{\rm s}/H_{\rm c2}\sim 0.35$ in our calculation 
becomes smaller than $H_{\rm s}/H_{\rm c2}\sim 0.5$ in ref. \citen{vorontsov2007}. 
$T_{\rm s}/T_{\rm c}$ is almost same in both calculations. 
The reason of the sign change is because of the $E$-dependence of DOS $N(E)$, 
as discussed in  ref. \citen{vorontsov2007}. 
Compared with $N(E)$ for ${\bf H}\parallel {\rm node}$, 
$N(E)$ for ${\bf H}\parallel {\rm antinode}$ is smaller at low $E$, but 
larger at higher $E$. 

Our new approach is to evaluate the contribution of paramagnetic effect,  
by comparing results in Figs. \ref{fig:T-C4mu2} and \ref{fig:T-C4mu0}. 
By the paramagnetic effect, the upper critical field is largely suppressed 
from $H_{\rm c2} \sim1.55$ (when $\mu=0$) to $H_{\rm c2}\sim 0.19$ 
(when $\mu=2$), and the sign change field is also largely suppressed from 
$H_{\rm s} \sim 0.54$  to $H_{\rm s} \sim 0.067$. 
Thus, in the scale $H/H_{\rm c2}$, the normalized sign-change field 
$H_{\rm s}/H_{\rm c2}$ keeps similar value, which is not seriously changed 
by the paramagnetic effect. 
We see also the difference at $H>H_{\rm s}$,
in Fig. \ref{fig:T-C4mu2},  $A_4$ monotonically increase on low temperature, 
but in Fig. \ref{fig:T-C4mu0},  $A_4$ decreases at low temperature 
after increase at high temperature. 
Thus, solid line of $H_{\rm s}$ almost horizontal at low $T$ in Fig. \ref{fig:T-C4mu2}(b), 
but the line of $H_{\rm s}$ increases on lowering $T$ in Fig. \ref{fig:T-C4mu0}(b).  

\begin{figure}[tb]
\begin{center}
\includegraphics[width=6.0cm]{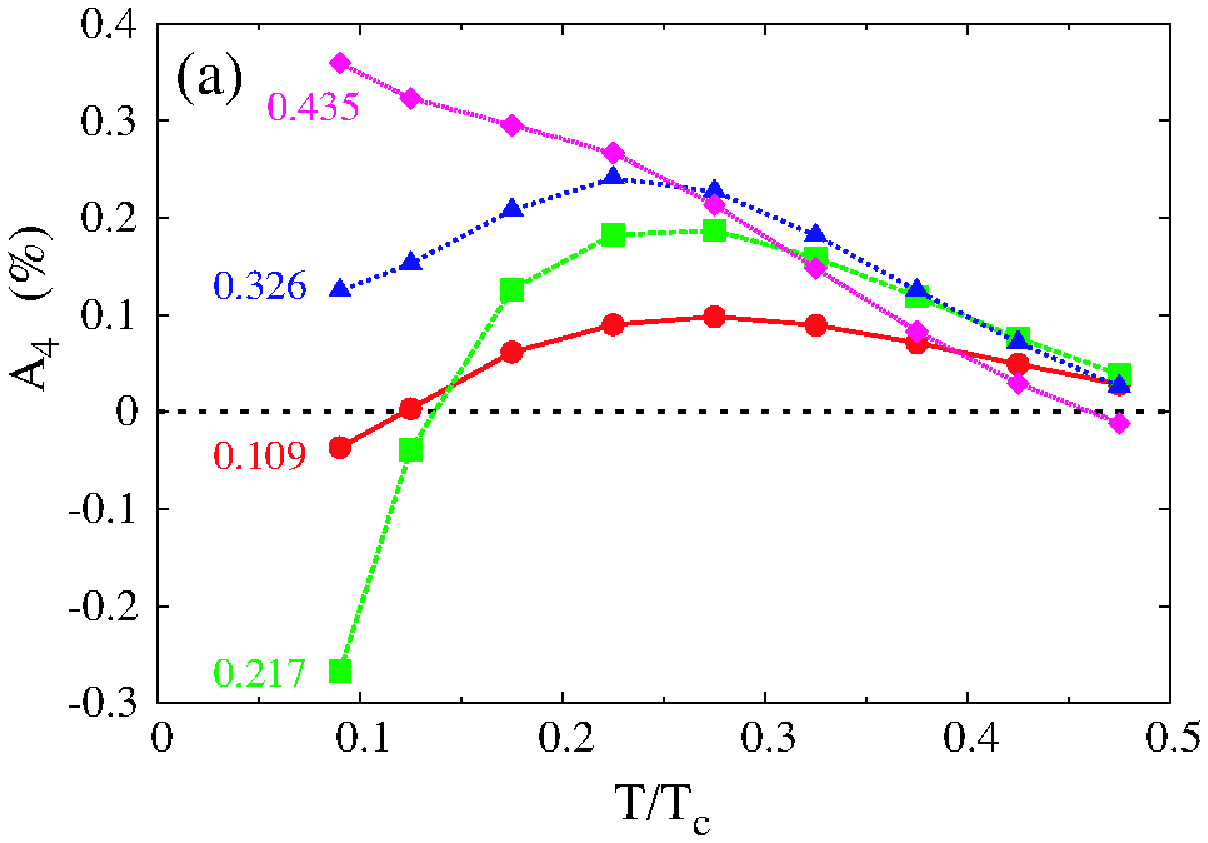}
\includegraphics[width=6.0cm]{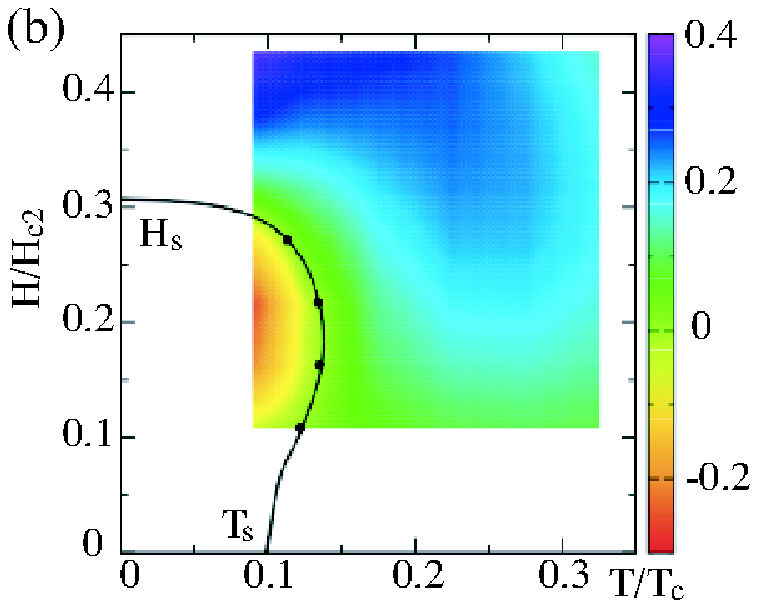}
\end{center}
\caption{(Color online) 
The same as Fig. \ref{fig:T-C4mu2}, 
but in the case when we include the contribution from anisotropic Fermi velocity by 
$\beta=0.5$ with strong paramagnetic effect $\mu=2$. 
In (a), $H/H_{\rm c2}=$0.109,  0.217, 0.326, and 0.435.
$H_{\rm c2} \sim 0.184$. 
}
\label{fig:T-C4mu2beta0.5}
\end{figure}

Lastly, we also evaluate the contribution of Fermi velocity anisotropy on $A_4$, 
considering finite $\beta$ in the definition of Fermi surface in eq. (\ref{eq:vf}).
Figure \ref{fig:T-C4mu2beta0.5} shows $A_4(T,H)$ for $\beta=0.5$ and $\mu=2$. 
This positive $\beta$ shifts $H_{\rm s}$ smaller to 
$H_{\rm s}/H_{\rm c2} \sim 0.3$  for $\beta=0.5$, 
from $H_{\rm s}/H_{\rm c2}=0.35$ for $\beta=0$ [Fig. \ref{fig:T-C4mu2}]. 
On the other hand, negative $\beta$ makes $H_{\rm s}$ larger. 
Therefore, anisotropic Fermi velocity affect on the sign-change field $H_{\rm s}$. 
However, from Figs. \ref{fig:T-C4mu2} - \ref{fig:T-C4mu2beta0.5},  
we see that the sign-change temperature $T_{\rm s}$ of $A_4$ is rather independent 
from the Fermi surface anisotropy and paramagnetic effect. 

Recently the sign-change of $A_4$ in specific heat oscillation 
was observe in ${\rm CeCoIn_5}$.\cite{an} 
There, $A_4$ shows the sign change at low $T$ and low $H$. 
This is qualitatively consistent to our results of calculations, 
and supports the $d_{x^2-y^2}$-wave pairing for superconductivity 
in ${\rm CeCoIn_5}$.  
Compared to the experimental data in ref. \citen{an}, 
amplitude of $A_4$ is smaller in our calculation. 
One of the reason is that in analysis of experimental data 
definition of $A_4$ is $C=C_0+C_{H}(1-A_4 \cos 4 \theta )$, 
while in our calculation $C_0=0$ in the clean limit. 
While $T_{\rm s}/T_{\rm c} \sim 0.1$ both for experiment and theory, 
$H_{\rm s}/H_{\rm c2} \sim 0.1$ in experimental data is smaller than 
the theoretical calculation. 
We showed that $H_{\rm s}/H_{\rm c2} $ can be changed by the Fermi velocity 
anisotropy, based on a simple $\beta$-model. 
Therefore, as a possibility, by theoretical estimate using realistic Fermi 
surface structure may improve quantitative accordance of $H_{\rm s}$ with 
experimental data.  

\section{Summary and Discussions}

We investigated the vortex state in $d_{x^2-y^2}$-wave pairing 
when magnetic field ${\bf H}$ is applied parallel to $ab$ plane,  
based on quantitative calculation by selfconsistent Eilenberger theory. 
Evaluating (1) stable vortex lattice structure, (2) the zero-energy LDOS 
in the vortex lattice state, and (3) temperature dependence of specific heat,    
we estimate the differences of vortex states  
for ${\bf H}\parallel{\rm node}$ and for ${\bf H}\parallel{\rm antinode}$.  

(1)
The transition of two possible vortex lattice configurations (type-A 
and type-B in Fig. \ref{fig:latticeD}) can occur as a function of 
temperature and magnetic field. 
There is a possibility that vortex lattice configuration for 
${\bf H}\parallel{\rm antinode}$ is different from that for 
${\bf H}\parallel{\rm node}$. 
This indicate that by rotation of magnetic field orientation within $ab$ plane,  
first order transition occurs between two vortex lattice configuration. 

(2) The spatial structure of zero-energy LDOS reflects Q2D Fermi surface 
structure. 
The tails of zero-energy LDOS prefer connecting with those of neighbor vortices  
in the $ab$ direction compared to those in $c$ direction. 
The dependence on the relative angle between the node-direction and magnetic field 
orientation gives minor contribution to broad extension of quasiparticles around 
vortex cores. 

(3) 
We evaluated 
magnetic field and temperature dependence of the amplitude and sign 
for specific heat oscillation by rotation of magnetic field orientation. 
The sign of the oscillation changes at low field and low temperature region, 
where the sign is consistent to the oscillation of evaluation by zero-energy DOS.
This sign-change behavior was recently observed in ${\rm CeCoIn_5}$.\cite{an}  
Our selfconsistent calculation without Pesch approximation gives 
qualitatively consistent results to those in previous work by Pesch approximation 
when the paramagnetic effect is absent.\cite{vorontsov2007} 
We extend this calculation to the case of strong paramagnetic effect 
such as ${\rm CeCoIn_5}$, 
and show the sign-change behavior of the specific heat oscillation 
is similar to that without paramagnetic effect, if the magnetic field 
is scaled by suppressed $H_{\rm c2}$.  
We also evaluate the contribution of anisotropic Fermi velocity, 
which shifts the sign-change field.

The experiments to observe the dependence on magnetic field orientation 
are useful method to identify the position of the node 
in the superconducting gap on the Fermi surface. 
In addition to specific heat oscillation by rotation of magnetic field orientation, 
we examine the dependence on the relative angle between the node and 
field orientation for stable vortex lattice configuration and spatial structure 
of LDOS, with and without paramagnetic effect.  
The theoretical calculations give valuable information to 
be compared with details of experimental data.

\section*{Acknowledgments} 
 
The authors are grateful for helpful discussions and communications with 
E.M. Forgan, J.S. White and T. Sakakibara.

\end{document}